\documentclass[12pt,preprint]{aastex}
\shorttitle{Frequency Of Close-In Jovian Planets Around M Dwarfs}
\shortauthors{Endl M., Cochran W.D., et al.}
\begin{document}

\title{Exploring The Frequency Of Close-In Jovian Planets Around M Dwarfs 
\footnote{Based on data collected with the Hobby-Eberly Telescope, which is operated
by McDonald Observatory on behalf of The
University of Texas at Austin, the Pennsylvania State University, Stanford University,
Ludwig-Maximilians-Universit\"at M\"unchen, and Georg-August-Universit\"at G\"ottingen.
Also based on observations collected at the European Southern Observatory, Chile (ESO Programmes 65.L-0428, 66.C-0446, 267.C-5700, 68.C-0415, 
69.C-0722, 70.C-0044, 71.C-0498, 072.C-0495, 173.C-0606). Additional data were obtained at the W.M.Keck Observatory, which is operated as a scientific partnership among the
California Institute of Technology, the University of California, and the National Aeronautics and Space Administration (NASA), 
as well as with the McDonald Observatory Harlan J. Smith 2.7~m telescope.}
}

\author{Michael Endl, William D. Cochran}
\affil{McDonald Observatory, The University of Texas at Austin, Austin, TX 78712}
\email{mike@astro.as.utexas.edu ; wdc@astro.as.utexas.edu}
\author{Martin K\"urster}
\affil{Max-Planck-Institut f\"ur Astronomie, K\"onigstuhl 17, Heidelberg, D-69117}
\email{kuerster@mpia.de}
\author{Diane B. Paulson}
\affil{NASA Goddard Space Flight Center, Planetary Systems Branch, Greenbelt, MD 20771}
\email{Diane.B.Paulson@gsfc.nasa.gov}
\author{Robert A. Wittenmyer, Phillip J. MacQueen, Robert G. Tull}
\affil{McDonald Observatory, The University of Texas at Austin, Austin, TX 78712}
\email{robw@astro.as.utexas.edu ; pjm@astro.as.utexas.edu ; rgt@astro.as.utexas.edu}

\begin{abstract}
We discuss our high precision radial velocity results of a sample of $90$ M~dwarfs 
observed with the Hobby-Eberly Telescope and the Harlan J. Smith 2.7~m Telescope
at McDonald Observatory, as well as the ESO VLT and the Keck I telescopes, within the
context of the overall frequency of Jupiter-mass planetary companions to main sequence stars.
None of the stars in our sample show variability indicative of a giant planet in a short 
period orbit, with $a\leq1$~AU. We estimate an upper limit of the frequency $f$ of close-in 
Jovian planets around M dwarfs
as $< 1.27\%$ (at the $1 \sigma$ confidence level). Furthermore, we determine the
efficiency of our survey to have noticed planets in circular orbits as $98\%$ for companions with
$m \sin i > 3.8~{\rm M}_{\rm Jup}$ and $a\leq0.7$~AU. For eccentric orbits ($e=0.6$) the
survey completeness is $95\%$ for all planets with $m \sin i > 3.5~{\rm M}_{\rm Jup}$ and $a\leq0.7$~AU.
Our results point toward a generally lower frequency of close-in Jovian planets for M dwarfs 
as compared to FGK-type stars. This is an important piece of information for our understanding of the 
process of planet formation as a function of stellar mass.
\end{abstract}


\section{Introduction}

Despite the stunning success of the radial velocity (RV) technique in finding extrasolar planets 
(e.g. Mayor \& Queloz 1995), which resulted in the discovery of more than $160$ planetary companions
over the past decade, our understanding of planet formation is far from complete. 
We are especially lacking a general overview of the frequency of planetary companions to stars throughout the 
entire HR-diagram. Doppler surveys have traditionally targeted bright solar type main sequence stars 
and it is no big surprise that most planets were found around G-type stars. But, is this entirely a result of
an observational bias, or is it a true effect which could lead to a better understanding of the underlying physics 
of planet formation? 

The majority of the stars in the solar neighborhood are M~dwarf stars with masses of $0.5~{\rm M}_{\odot}$ or less
(Henry~1998). In order to determine the overall galactic population of planets it is important not to ``overlook''
the faint and low mass regime of the HR-diagram and to control observational biases in order to arrive
at statistically meaningful results. 
Because of the intrinsic faintness of M dwarfs it is generally more difficult 
and time consuming to achieve the high quality RV data for the detection of planetary companions. 
This led to the situation that M dwarfs constitute only small subsets in the target samples of most
precision Doppler surveys.

So far, we know of only one M dwarf, GJ~876 (M4~V, M=$0.3~{\rm M}_{\odot}$), to harbor a planetary system
with Jupiter-mass companions (Delfosse et al.~1998; Marcy et al.~1998; Marcy et al.~2001; 
Benedict et al.~2002). There is the possibility that the giant planet detected by microlensing 
(Bond et al.~2004) also orbits an M~dwarf, but the exact spectral type of the primary lens has 
not been determined yet. Butler et al.~(2004) announced the discovery of a short periodic Neptune mass 
planet around the M dwarf GJ~436 and Rivera et al.~(2005) presented evidence for an extremely low
mass ($M\approx7.5~{\rm M}_{\rm Earth}$) third planetary companion to GJ~876. Bonfils et al.~(2005b) 
reported the detection of a Neptune mass companion to the southern M3~V star GJ~581.

Endl et al.~(2003) described the dedicated M dwarf survey at the Hobby-Eberly Telescope (HET)
which targets exclusively M dwarfs and presented the data of the first year of the survey. 
This paper now contains three years of RV results from the on-going HET program with the addition 
of five years of the M~dwarf RV results from our planet search program at the ESO Very large Telescope (VLT), as
well as data from the McDonald 2.7~m telescope program and from the Keck Hyades survey. 
We discuss these results and their implications on the total frequency of detectable giant planets 
along the main sequence.  
   
\section{M dwarf radial velocity results}

The precise M dwarf RV results we present here originate from four different Doppler surveys: the 
majority of the data stem from our dedicated M dwarf survey carried out at the HET (Endl et al.~2003),
20 targets in the southern hemisphere are part of our VLT program (K\"urster et al.~2003 ; K\"urster \& Endl~2004) 
while 6 M~dwarfs (GJ~15~A, GJ~411, GJ~412~A, GJ~671, GJ~725~A \& B) were monitored as part of our 
long term survey using the Harlan J. Smith (HJS) 2.7~m telescope at McDonald 
(e.g. Cochran et al.~1997 ; Hatzes et al.~2003 ; Endl et al.~2004) and 
15 M~dwarfs were part of the Keck I HIRES survey of the Hyades cluster (Cochran, Hatzes, 
Paulson~2002 ; Paulson, Cochran, Hatzes~2004). 

Targets are selected based on the Gliese catalog of nearby stars (Gliese \& Jahrei\ss~1991)
and the $Hipparcos$ catalog (Perryman~1997).
We choose nearby M dwarfs which are brighter than $V=12$ in order to obtain high resolution spectra with sufficient
signal-to-noise ratios for precise RV measurements (with the exception of a few Hyades targets). 
In general we excluded M stars from our survey which 
show strong coronal X-ray emission using the ROSAT All-Sky-Survey data (H\"unsch et al.~1999) to minimize additional
RV noise due to stellar activity (exceptions again are Hyades stars and Proxima Cen). 
Unlikely planet hosts like short periodic binaries are also not included in our
survey (with the exception of GJ~623). And for the Hyades survey the brightest M dwarfs in the cluster were selected.
Because GJ~876 has already known planetary companions we did not include this M dwarf in our program.

Table 1 lists all $90$ M dwarfs along with their spectral type, visual magnitude, distance (based
on their $Hipparcos$ parallax), 
number of measurements, total RV rms scatter, mean measurement uncertainty and the duration of monitoring. 
The M~dwarfs included in this study have a mean magnitude of $V=10.57$~mag, the brighest target has 
a $V$ magnitude of $7.48$ and the faintest star has $V=12.98$. 
On average we obtained $16$ measurements per target. The mean time coverage of the targets is $1114$~days, with
$333$ days as the shortest monitoring time span and $2980$ days as the longest.
$1114$~days corresponds to the period of an orbit with
a semi-major axis of $\approx1.7$~AU for a $0.5~{\rm M}_{\odot}$ star.

The $82$ M dwarfs of the sample which have a $Hipparcos$ parallax measurement
are located at a mean distance of $16.9$~pc. The closest star is GJ 551 (Proxima Cen) with 
$d=1.29$~pc and the most distant target with a $Hipparcos$ parallax is at a distance of
$d=58.1$~pc (HIP~16548).

\begin{figure} 
\includegraphics[angle=270,scale=0.6]{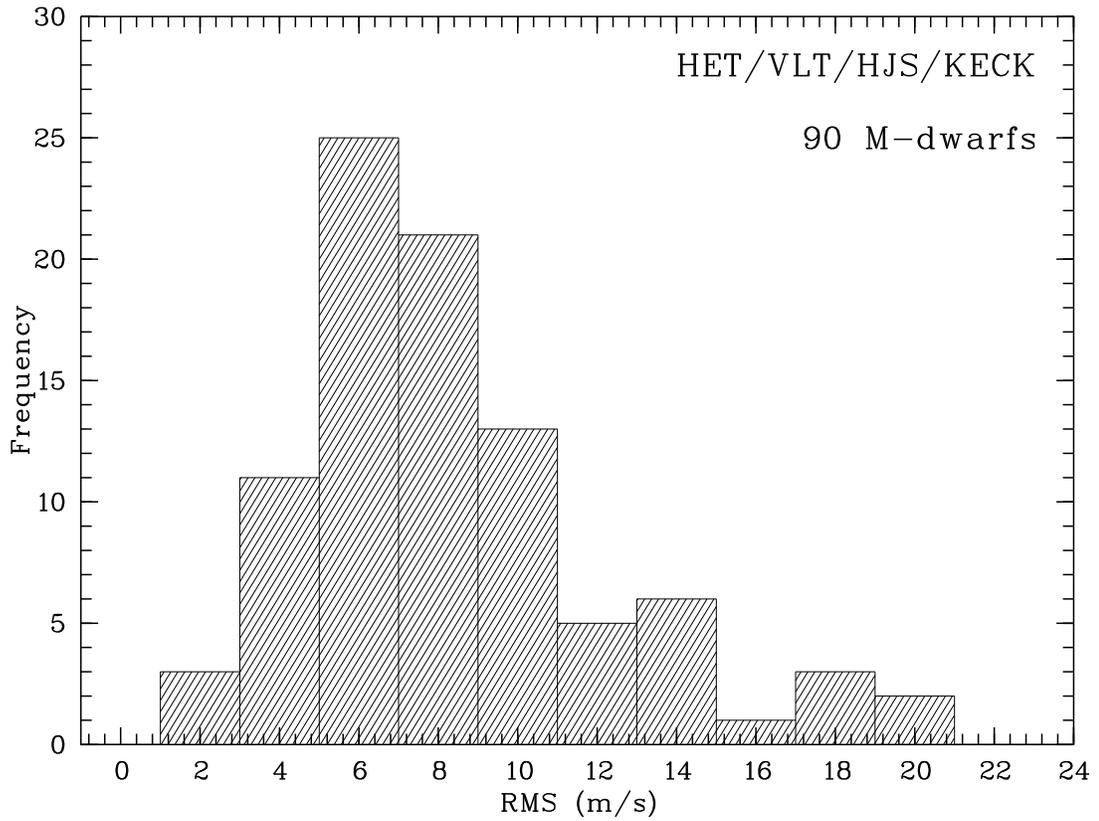}
\caption{Histogram of the total RMS scatter for the $90$ M-dwarfs in the sample
observed with the HET, VLT, HJS and Keck telescopes.
The mean value of this distribution is $8.3~{\rm m\,s}^{-1}$ 
with a $\sigma$ of $3.9~{\rm m\,s}^{-1}$.}
 \label{histo}
\end{figure}

Fig.~\ref{histo} shows the histogram of the total rms scatter of the RV data.
The main peak of the distribution is around $6.0~{\rm m\,s}^{-1}$. 
The mean RV scatter is $8.3~{\rm m\,s}^{-1}$ with a $\sigma$ of $3.9~{\rm m\,s}^{-1}$. 
There is a small secondary peak containing $6$ stars with rms$>15.0~{\rm m\,s}^{-1}$:  
GJ~436, which has a low mass planetary companion in a short-period orbit (Butler et al.~2004), and $5$ young and active 
Hyades M dwarfs: BD+07~499, HD~286554, HIP~16548, vA~115 \& vA~502. 
  
A more detailed description of the RV results for GJ~1, GJ~15~A (binary), GJ~270 (binary), 
GJ~310 (binary), GJ~551, GJ~623 (binary), GJ~699 (= Barnard's star), GJ~725~A\&B (binary), and GJ~846 (erratum) is given in the Appendix.

None of the M dwarfs surveyed by our programs reveal an increased RV scatter which can be attributed
to Keplerian reflex motion caused by a Jovian planetary companion with $P<T_{\rm Survey}$.

\section{M dwarf planet frequency}

With this null result in hand, what kind of conclusions can we draw about the frequency of giant planets
orbiting M dwarfs? We follow the procedure outlined in the Appendix of Burgasser et al.~(2003) and 
McCarthy \& Zuckerman (2004) and estimate the companion frequency $f$ by using the binominal distribution:

\begin{equation}
P_{d}(f) = f^{d}(1-f)^{N-d} \frac {N!} {(N-d)!d!} .
\end{equation}

$P_{d}(f)$ is the probability that an ideal survey of $N$ targets will yield a detection rate $d$, 
for a true frequency of companions $f$.

\begin{figure}
\includegraphics[angle=270,scale=0.6]{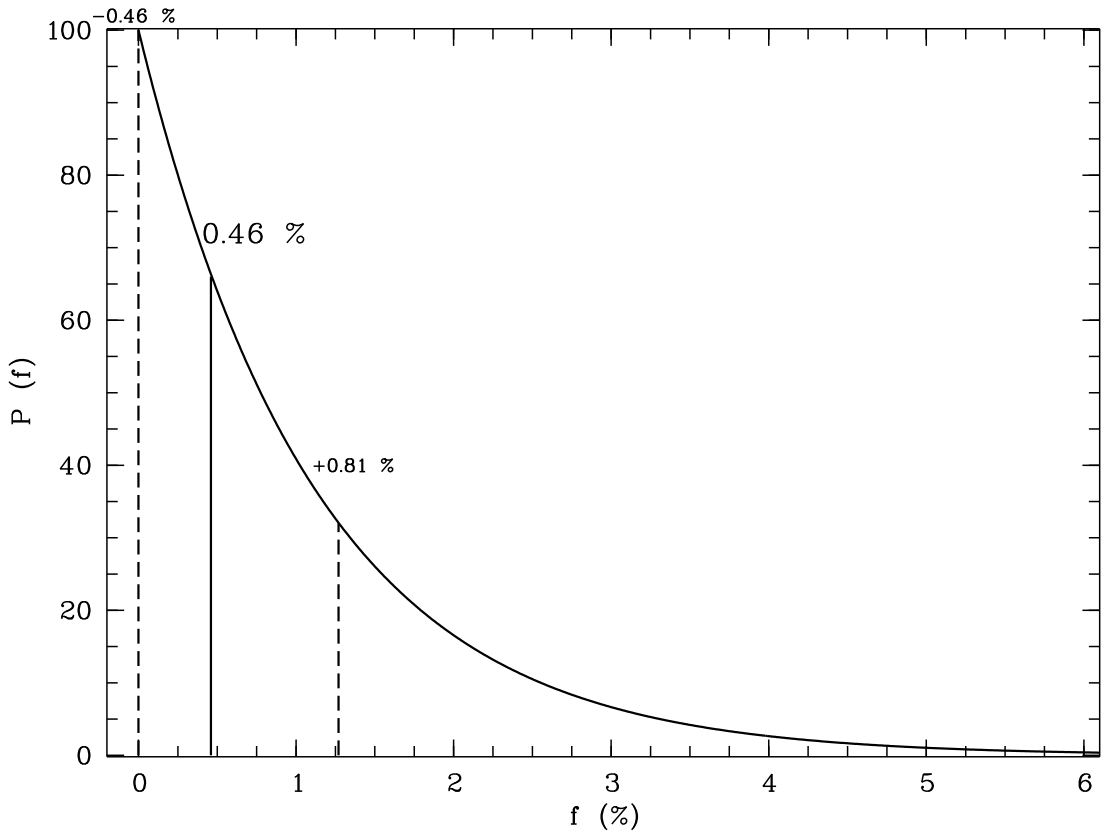} 
\caption{Probability function $P(f)$ for the true companion frequency $f$ based on all our M dwarf data 
(HET, VLT, HJS, and KECK: $N=89$ stars) and $d=0$ detections. We find $f=0.46_{-0.46}^{+0.81}\%$. 
The dashed lines delimit the area of $68\%$ integrated probability ($\approx 1\sigma$ Gaussian error).}
 \label{prob}
\end{figure}

For the estimation of the M dwarf planet frequency we remove GJ~623 from the sample, as it is
a short period binary with an eccentric orbit. 
Using only the remaining M dwarf sample (number of stars $N=89$ and detections $d=0$) we find 
$f=0.46_{-0.46}^{+0.81}$\% (see Fig.~\ref{prob}). The error bar denotes the area of $68$\% integrated
probability (i.e. $1\sigma$ confidence). At this confidence level we thus derive an upper limit for $f$ of
$1.27$\%.
 
\subsection{Survey completeness}

Of course, no real survey is an ideal survey. We estimate the completeness of our M dwarf survey 
by using numerical simulations. We start with the null hypothesis that the observed scatter represents
the distribution of our measurement errors (i.e. that no additional signal is buried in the data). 
For each star we then add Keplerian signals to our data (at the times of observation) 
and compare the resulting new rms scatter with the
originally observed value. If the F-test shows a probability of less than $99\%$ that these two variances are drawn 
from two different samples, the planetary companion corresponding to the input signal is declared as 
``missed'' by our program. For each period and amplitude we compute Keplerian signals at $10$ different orbital 
phases and for eccentric orbits for each phase at $10$ different periastron angles. This results in $100$ simulated
planets per period, amplitude and target. The amplitude of the input signal is increased until a certain
overall threshold (e.g. less than $10\%$ or $5\%$ of all planets are missed) is reached.

\begin{figure}
\includegraphics[angle=270,scale=0.6]{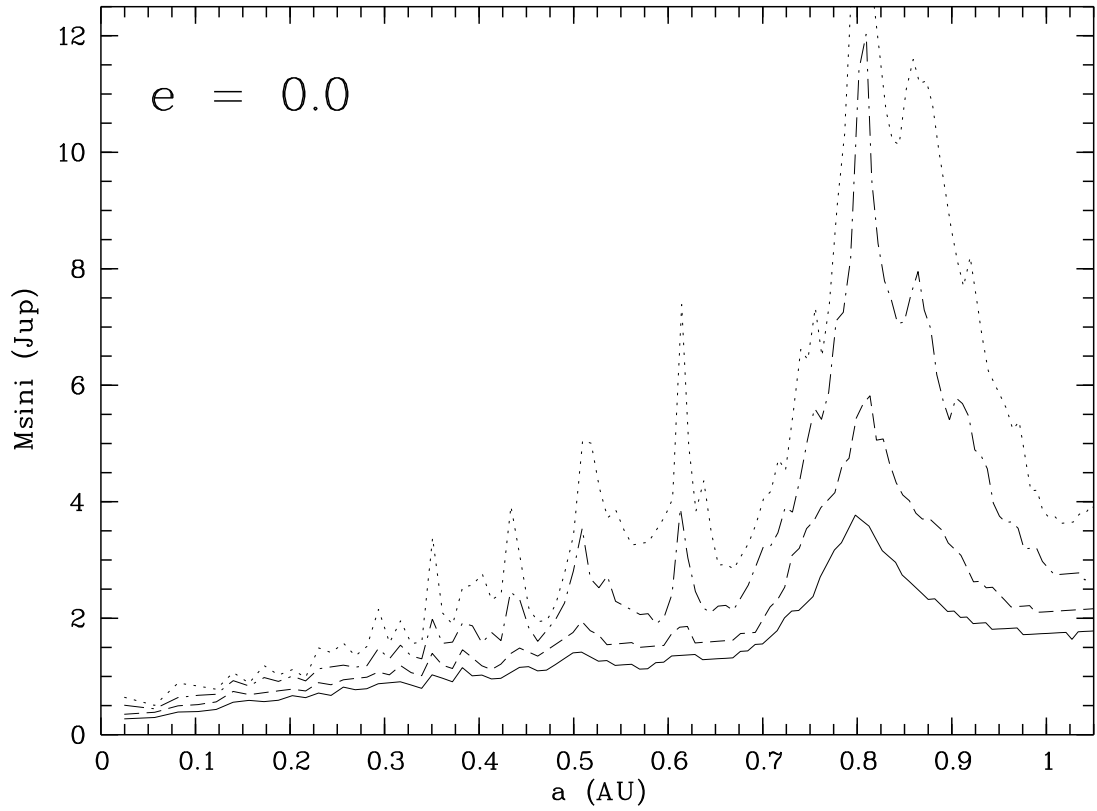}
\caption{Estimated survey completeness for planets in circular orbits. The lines represent
$90\%$ success rate (solid line), $95\%$ (dashed line), $98\%$ (dash-dotted line) and 
$99\%$ (dotted line).}
 \label{sens1}
\end{figure}

Fig.~\ref{sens1} displays the survey sensitivity determined by this method for the case of circular orbits with semi-major 
axes $a = 0.025$ to $1.0$~AU. The limits shown in the figure are for $90\%$, $95\%$, $98\%$ and $99\%$ survey 
completeness. 
To transform the amplitude information into a mass value we adopt a primary mass of $0.5~{\rm M}_{\odot}$. Note
that the mass limits for the planets are conservative upper limits because for less massive stars these companion mass 
values would be lower. For less massive M dwarfs this diagram would also move 
inwards (in terms of orbital semi-major axis), because companions around less massive stars orbit closer to the primary
at a given orbital period. 

Based on these results we find that we have a $95\%$ efficiency to notice all planets with 
$m \sin i > 2.3~{\rm M}_{\rm Jup}$ at $a \leq 0.7$~AU and of $98\%$ for all $m \sin i > 3.9~{\rm M}_{\rm Jup}$ 
companions at these orbital separations. With the exception of a sharp spike up to $7.4~{\rm M}_{\rm Jup}$ 
at $a=0.61$~AU we are $99\%$ complete for all planets with $m \sin i > 5.1~{\rm M}_{\rm Jup}$ at $a \leq 0.7$~AU.
For periods close to 1 year ($a \approx 0.8$~AU for a $0.5~{\rm M}_{\odot}$ star) 
the efficiency drops rapidly due to the window function: 
$90\%$ of all planets with $m \sin i > 4.0~{\rm M}_{\rm Jup}$ and $95\%$ of all
$m \sin i > 6.6~{\rm M}_{\rm Jup}$ companions.

\begin{figure}[t]
\includegraphics[angle=270,scale=0.6]{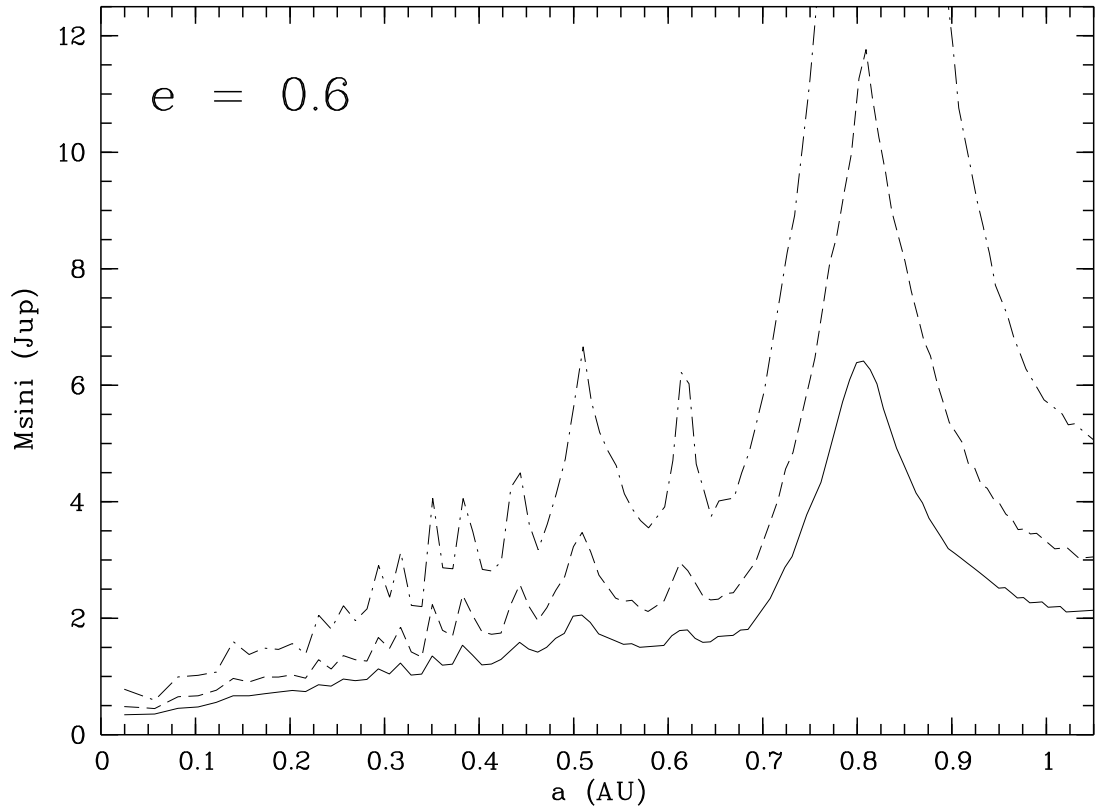}
\caption{Estimated survey completeness for planets in eccentric orbits with $e=0.6$. 
The lines represent $90\%$ success rate (solid line), $95\%$ (dashed line) and $98\%$ (dash-dotted line).} 
 \label{sens2}
\end{figure}

Orbital eccentricity decreases the survey sensitivity further, because a Doppler survey can easily miss 
critical orbital phases of eccentric orbits due to sparse sampling and the star can thus appear to be RV constant. 
We repeated the simulations with an orbital eccentricity of $e=0.6$ 
(which would include the majority of known extrasolar planets) and the results are displayed in Fig.~\ref{sens2}.
For orbital separations of $a\leq0.7$~AU we are $95\%$ complete for planets with $m \sin i > 3.4~{\rm M}_{\rm Jup}$
and $98\%$ for $m \sin i > 6.7~{\rm M}_{\rm Jup}$. Again, the survey efficiency drops rapidly for periods close
to 1 year. 

\section{Discussion}

Marcy et al. (2005) find a frequency of $1.2\pm0.2\%$ of ``hot Jupiters'' with $a<0.1$~AU around
FGK-type stars and according to their Fig.2 a frequency of $2.5\pm0.4\%$ of planets
with $a<1$~AU. While the frequency of ''hot Jupiters'' is still consistent
with our results, it appears that there is a difference emerging between the detection rate of
Jovian planets with $a<1$~AU around FGK-type stars and M dwarfs.
Also Butler et al.~(2004) noted that the combined results from all 
radial velocity planet search programs, which include M dwarfs in their target samples, point 
toward an upper limit for $f$ of $<0.5\%$.
Lineweaver \& Grether (2003) presented an analysis and extrapolation of the planet frequency for
FGK-type stars currently monitored by Doppler surveys. They find that the fraction of detected planets increases
from $\approx3.9\pm1\%$ for a survey duration of 2 years to $5.5\pm1.5\%$ for a 4 year survey. 
Again, both values are higher than the upper limit we find. However, our upper limit of $1.27\%$ for M dwarf planets 
is only valid in the planet mass-separation range, to which our program is most sensitive (with a $>98-99\%$ completeness,
see Fig.~\ref{sens1} and Fig.~\ref{sens2}). But, also FGK-star surveys are not ideal surveys and they
miss a small fraction of planets. 

Gaudi et al.~(2002) also finds a low frequency for Jupiter-mass companions around M dwarfs in the galactic
bulge based on the results of the PLANET microlensing survey.
However, one has to bear in mind that the PLANET survey samples a different region
of our galaxy and a comparison with the solar neighborhood could be inadequate.

A general observational bias as explanation for a lower M dwarf Jovian planet frequency 
becomes increasingly unlikely but is not completely ruled out. Doppler surveys of FGK-type stars
usually have a higher RV precision than M dwarf programs. The bias introduced by a somewhat lower
RV precision for M dwarfs is partly compensated by the fact that
the RV amplitudes induced by giant planets around M dwarfs are larger because of the lower mass
of the host star and thus easier to detect. Using large aperture telescopes allows us
to overcome their intrinsic faintness and to obtain high quality RV data also for these late spectral types.
Continuation, improvement and expansion of current M dwarf Doppler surveys is necessary to allow a better determination of
the planet frequency around these stars and to perform a more detailed comparison with the results from
the FGK-star surveys. 

Another issue which could still lead to an unintentional observational bias in our M dwarf sample is stellar metallicity. 
In recent years, it has became more and more obvious that the frequency of detectable planets is a function 
of the metal content of the stars included in Doppler surveys (Fischer \& Valenti~2005 ; Santos, Israelian, \& Mayor~2004). 
Metal poor stars appear to harbor fewer planets detectable by the RV technique (close in massive planets) than metal 
rich stars. This could mean that the formation mechanism is somehow linked to the metal content in the 
protoplanetary disk (at least for the stars included in these surveys). 
Is it possible that we mostly targeted metal poor M dwarfs and that this is the reason for the observed low planet 
frequency?
Our HET and VLT samples are biased toward inactive and thus presumably older M dwarfs.
Thus, it is conceivable that the majority of these M dwarfs are metal poor. A survey including also more active
M dwarfs as well as the determination of the metallicities of the targets can solve this issue.
Unfortunately, there are no large scale spectroscopic surveys to determine precise metallicities of M dwarfs, although there are efforts 
under way to find suitable techniques (e.g. Valenti et al.~1998). Woolf \& Wallerstein~(2005) 
presented [Fe/H] measurements for 35 M and K dwarfs. Three M dwarfs from our sample, GJ~411, GJ~412~A, and GJ~687 are included
in their study. They find that GJ~411 and GJ~412~A are metal poor (${\rm [Fe/H]}\approx-0.4$), while GJ~687 is metal rich
(${\rm [Fe/H]}\approx0.15$). Iron abundances for three more stars (GJ~15~A, GJ~109, GJ~849) of our sample are estimated by 
Bonfils et al.~(2005a). These authors used visual binaries with M dwarf secondaries to calibrate a photometric
method to derive M dwarf metallicities. GJ~15~A (${\rm [Fe/H]}\approx-0.45$) and GJ~109 (${\rm [Fe/H]}\approx-0.2$) 
turn out to be metal poor stars, while GJ~849 (${\rm [Fe/H]}\approx0.14$) has a higher than solar iron abundance.
So far, four out of six target stars with [Fe/H] measurements thus turn out to be metal poor stars.  
A detailed determination of the metallicity of our sample stars 
using the method of Bean et al.~(2005) is currently in progress.
The Hyades M dwarfs are of course young and active stars. They should have a similar 
metal content as the cluster mean of [Fe/H] =$0.13\pm0.01$ (Paulson, Sneden, Cochran~2003). But the small 
number of M dwarfs included in the overall Hyades sample 
prevents a meaningful comparison with the planet frequency of earlier type stars in that metallicity range.

How do our observations agree with current models for planet formation? 
Laughlin, Bodenheimer \& Adams~(2004) explored the formation of gas giants around M dwarfs
within the framework of the core accretion model (e.g. Pollack et al.~1996). These authors conclude that this model
has severe problems in forming Jupiter-class planets in the less massive protoplanetary disks of M dwarfs.
Also, Ida \& Lin~(2005) show that close-in Jovian planets should be relatively rare around M dwarfs. 
Our data confirm so far their theoretical prediction that M dwarfs should harbor fewer gas giant planets. 
On the other
hand, Boss~(2006) shows that the gravitational instability model has a higher efficiency in forming giant planets in less massive M dwarf disks. 
GJ~876 remains the only unambiguous case of an M dwarf orbited
by Jupiter-mass planetary companions despite the accumulation of high quality RV data for more M dwarfs
over the past years. So what makes GJ~876 so special? At the moment we can only speculate that 
GJ~876 had either a more massive disk than other M dwarfs, albeit for unknown reasons, or these planets indeed
formed by gravitational instability. As suggested by Boss~(2006), M dwarfs could be a used as a test ground for 
competing giant planet formation models.

In the case that orbital migration (e.g. Lin et al.~1996) results in the observed small semi-major axes for many of the planets detected by
Doppler surveys, we can pose the question of whether this mechanism might be less efficient for M dwarfs. This would
explain the current null detections (again with the exception of GJ~876) because high quality RV data for
most M dwarfs do not have sufficient time coverage to find (or exclude) Jupiter-type companions in 
long periodic orbits. 
In our sample we do observe RV trends indicating more distant companions, but most of them are quite
large and most likely caused by stellar companions. Highly precise astrometric studies 
using data from the upcoming Space Interferometry Mission (SIM) 
should allow the detection of planets at large orbital separations around nearby M dwarfs.  
Firm upper limits from astrometry have been placed by
Benedict et al.~(1999) on the masses of planetary companions in the period range from
$60$ to $600$ days, orbiting Barnard's star and Proxima Centauri, using HST Fine Guidance Sensor data. 
Benedict et al.~(2002) succeeded in detecting the astrometric perturbation caused by the outermost
planet in the GJ~876 system, and derived a mass for the companion of $1.89\pm0.34~{\rm M}_{\rm Jup}$.

Despite their lack of close-in Jovian planets, M dwarfs remain attractive targets for current Doppler surveys. 
Because of their low primary masses the RV amplitude induced by an orbiting companion is higher than
for F,G, or K type stars. 
The discovery of planets with extremely low masses (a few ${\rm M}_{\rm Earth}$) 
using precise radial velocity measurements is feasible in the M dwarf regime (e.g. K\"urster et al. 2003). 
Model calculations (e.g. Wetherill~1996 ; Laughlin et al.~2004) also
show no difficulties in forming low mass planets via planetesimal accretion in disks around M dwarfs.
Ida \& Lin (2005) even predict a higher frequency of icy giant planets with masses comparable to
Neptune in short periodic orbits for M dwarfs than for G type stars.   
The lowest mass extrasolar planet our group has found so far is the fourth companion in 
the $\rho^{1}$~Cnc (G8~V) system (McArthur et al.~2004). 
The RV semi-amplitude induced by this planet is only $6.7~{\rm m\,s}^{-1}$ and hence a difficult signal to detect.   
If the same planet (${\rm P}=2.81~{\rm days}, m \sin i = 14.2~{\rm M}_{\rm Earth}$) were to orbit a high mass M dwarf 
($0.5~{\rm M}_{\odot}$) the RV semi-amplitude would increase to $10.2~{\rm m\,s}^{-1}$ and for Proxima
Centauri (M5~V, $0.12~{\rm M}_{\odot}$) the signal would be $26.3~{\rm m\,s}^{-1}$.   
The discoveries of the short-period Neptunes by Butler et al.~(2004) and Bonfils et al.~(2005b) and of an additional 
planet with an extremely low mass of $m \approx 7.5~{\rm M}_{\rm Earth}$ in the GJ~876 system (Rivera et al.~2005)
further support the notion that M dwarf stars remain a fruitful and interesting hunting ground
for high precision radial velocity surveys. In the future high resolution spectrometers
working in the near infrared will be the ideal tools for a more thorough 
exploration of the red (and low mass) part of the main sequence.

The NASA Kepler mission (Borucki et al.~2003), planned for launch in late
2008, should give much better insight into the true frequency of
short-period planets ($P < 1$ year) around M~dwarfs, provided a
sufficiently large sample of M dwarfs is included in the Kepler target
list.  M~dwarfs are particularly good targets for large photometric transit
surveys such as Kepler because their significantly smaller radius than
solar-type stars results in a much larger photometric signal for a given
planet size.

\acknowledgements
We thank the anonymous referee for many suggestions which helped to improve the manuscript.
We are grateful to the McDonald Observatory Time Allocation Committee and the ESO OPC for 
generous allocation of observing time. The help and support of the HET staff and especially of the resident 
astronomers: Matthew Shetrone, Brian Roman, Steven Odewahn and Jeff Mader were crucial for this project.    
This material is based upon work supported by
the National Aeronautics and Space Administration under Grant NNG04G141G. 
DBP is currently a National Research Council fellow working at NASA's Goddard Space Flight Center.
This research has made use of the SIMBAD database, operated at CDS, Strasbourg, France.

\section{APPENDIX}

\subsection{Individual results}

\subsubsection{Secular RV accelerations: GJ~1, GJ~411, GJ~551 and GJ~699}

For these 4 nearby M dwarfs we subtract from our data the expected secular acceleration of the RV caused by their 
proximity and/or large space motion. A detailed discussion of this effect and its measurement using our data for Barnard's star 
(GJ~699) is given by K\"urster et al.~(2003). The value of the secular accelerations for GJ~1 is 
$3.7~{\rm m\,s}^{-1}\,{\rm yr}^{-1}$,
for GJ~411: $1.35~{\rm m\,s}^{-1}\,{\rm yr}^{-1}$, for GJ~551 (Proxima Cen): $0.45~{\rm m\,s}^{-1}\,{\rm yr}^{-1}$, 
and for GJ~699: $4.5~{\rm m\,s}^{-1}\,{\rm yr}^{-1}$.     
 
\subsubsection{GJ~15~A, GJ~270, GJ~310 \& GJ~725~A\&B}
 
GJ~15~A (= HIP 1475 = HD~1326~A) is the primary of an M dwarf binary system with an angular separation to component
B of $36$ arcseconds. We detect a linear RV trend of $+1.45\pm0.40~{\rm m\,s}^{-1}\,{\rm yr}^{-1}$ ($\chi^{2}_{\rm red}=1.47, {\rm dof}=27$), 
which might be part of the long period orbit of the primary in this system. However, part of this RV trend is also the
expected secular acceleration of $0.7~{\rm m\,s}^{-1}\,{\rm yr}^{-1}$.  

The M0~V star GJ~270 (= HIP 35495 = G~87-33 = BD+33~1505) also exhibits a linear RV acceleration.
We fit a trend of $+171.4\pm2.2~{\rm m\,s}^{-1}\,{\rm yr}^{-1}$ with a $\chi^{2}_{\rm red}$ of $1.42$ (${\rm dof}=26$).
The residual rms scatter of the RV measurements around this linear RV trend is $12.0~{\rm m\,s}^{-1}$.
This trend is presumably caused by a previously unknown stellar
companion in a long periodic orbit. The $Hipparcos$ data for this star do not detect any astrometric
perturbation which also points toward a long period of the binary.
 
For GJ~310 (= HIP 42220 = G~234-38 = BD+67~552) we have almost exactly the same situation as for GJ~270.
We detect a linear RV trend of similar magnitude. But in this case a stellar secondary
with a long period of $\approx~24~{\rm yrs}$ (Heintz \& Cantor~1994) is already known. This companion
is very likely the cause of the observed RV trend.
We fit a linear RV trend of $+199.0\pm0.6~{\rm m\,s}^{-1}\,{\rm yr}^{-1}$
with $\chi^{2}_{\rm red}=10.6$ (${\rm dof}=36$).
Because the angular separation of the two components is only $0.55$~arcsecs, we
cannot rule out that spectral contamination by the secondary is the cause of the high $\chi^{2}_{\rm red}$ value and
large residual scatter of $13.7~{\rm m\,s}^{-1}$.

GJ~725 is a known binary consisting of 2 M dwarfs at an angular separation of 13.3 arcseconds (projected minimum
separation is $\approx 47$~AU). 
For component
A we find a linear RV trend of $+6.99\pm0.86~{\rm m\,s}^{-1}\,{\rm yr}^{-1}$ ($\chi^{2}_{\rm red}=0.9, {\rm dof}=21$) and
for component B of $-4.99\pm1.12~{\rm m\,s}^{-1}\,{\rm yr}^{-1}$ ($\chi^{2}_{\rm red}=0.67, {\rm dof}=18$). 
Because the trends have opposite signs it is very probable that in both cases we see a small section of the binary orbit. 

\subsubsection{GJ~623}
 
\begin{figure}
\includegraphics[angle=270,scale=0.6]{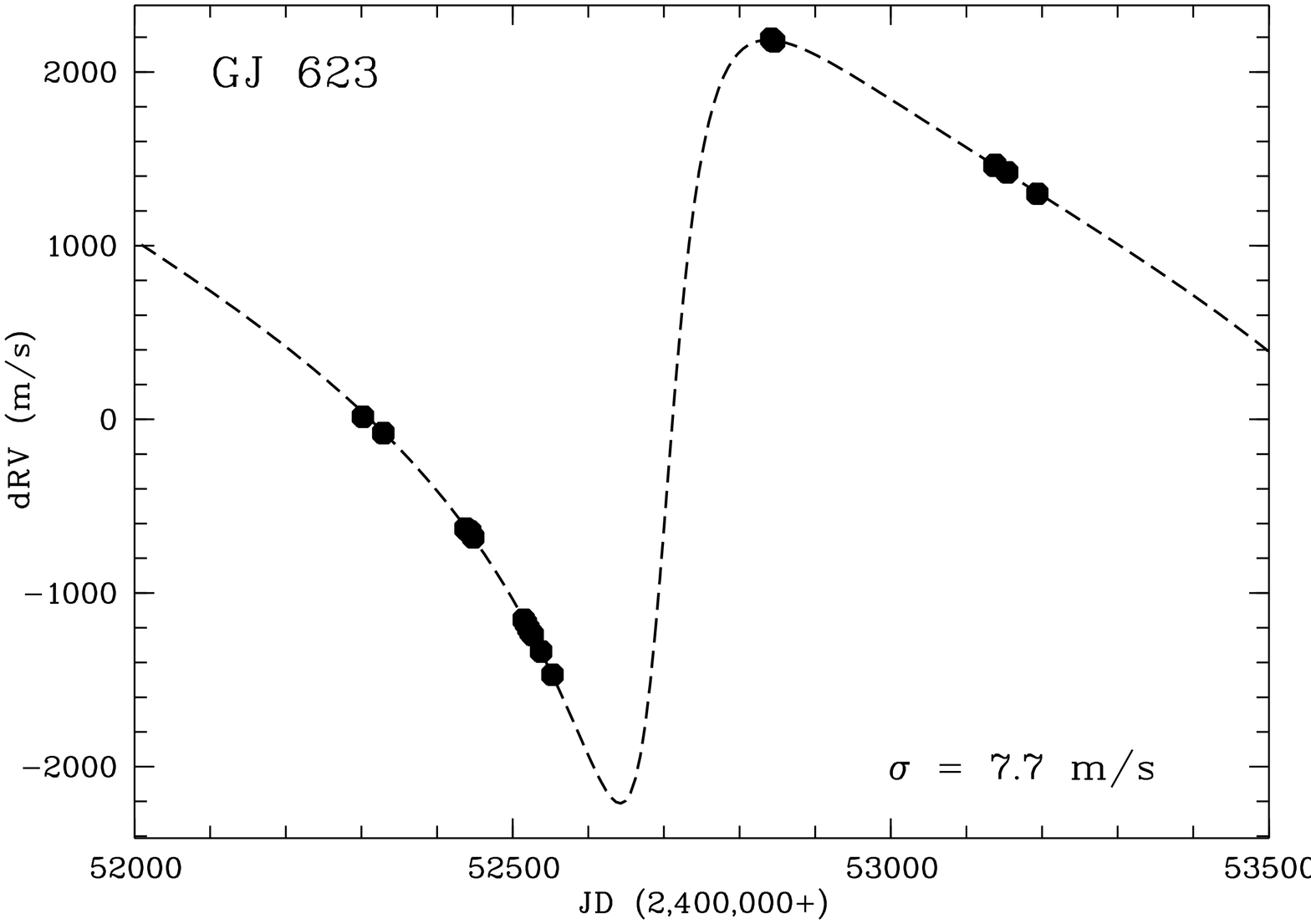} 
\caption{HET RV data of GJ 623 (filled circles) with the best-fit Keplerian orbital solution (dashed line) for
the binary over-plotted.}
 \label{gj623}
\end{figure}

For the known binary GJ~623 (Lippincott \& Borgman~1978) we find an orbital solution yielding the
following parameters: ${\rm P}=1291.8\pm15.9~{\rm days},~{\rm T}_{\rm periastron}=2451402.8\pm29.7,
~{\rm K}=2198.0\pm14.7~{\rm m\,s}^{-1},
~e=0.618\pm0.016,~\omega=245.7\pm1.2$, close to the latest published values by Nidever et al.~(2002).
The residual rms scatter around this orbit is $\sigma=7.7~{\rm m\,s}^{-1}$ ($\chi^{2}_{\rm red}=0.6$, ${\rm dof}=16$).
The HET data along with the Keplerian orbital solution are displayed in Fig.\ref{gj623}. 
The high eccentricity and short period
of the binary orbit makes GJ~623~A an unlikely host star for a close-in planetary companion (but
see Konacki 2005 for an example of a hot Jupiter in a tight binary system).
Assuming a stellar mass of $0.3\pm0.1~{\rm M}_{\odot}$ we derive a minimum mass value for the secondary
of $m \sin i = 41.5\pm9.0~{\rm M}_{\rm Jup}$. A combination of these RV data with HST/FGS astrometry and other 
RV data sets will allow a further refinement of the orbit and will result in an accurate mass 
for the secondary companion (Benedict et al., in prep.).
 
\subsubsection{Erratum: GJ~864}
 
In Endl et al. (2003) we discussed our results for GJ~864, which showed a large linear RV
trend of $\approx-2.0~~{\rm km\,s^{-1}\,yr^{-1}}$,
possibly due to a previously unknown stellar companion. Unfortunately, shortly after publication
we found an error in the coordinate entry for this target which introduced a systematic error into the
correction to the barycenter of the solar system. After correction of this mistake we find a much
smaller linear RV trend of $-32.3\pm2.4~{\rm m\,s^{-1}\,yr^{-1}}$ ($\chi^{2}_{\rm red}=0.97$, ${\rm dof}=25$). 
The residual scatter around this
trend is $10.4~{\rm m\,s}^{-1}$ almost identical to our average uncertainty of $10.7~{\rm m\,s}^{-1}$.
The classification of GJ~864 as RV constant star using the less precise CORAVEL data by Tokovinin~(1992)
is correct, considering the much shallower trend now. The companion causing this linear trend has probably a period
far exceeding our $999$ days of monitoring.

\clearpage
\begin{deluxetable}{lrlrrrrrrr}
\tabletypesize{\scriptsize}
\tablecaption{
The sample of $90$ M~dwarfs surveyed with the
HET, ESO VLT, HJS 2.7~m and Keck telescopes. Spectral type, visual magnitude ($V$), distance ($d$), number of
measurements ($N$), total RV scatter ($\sigma$), average internal measurement error ($\overline{\sigma}_{\rm int}$) 
and duration of monitoring ($\Delta$T) are given.\label{vels}}
\tablewidth{0pt}
\tablehead{ \colhead{Star} & \colhead{HIP} & \colhead{Sp.T.} &
\colhead{V} & \colhead{d} & \colhead{N} & 
\colhead{$\sigma$} &
\colhead{$\overline{\sigma}_{\rm int}$} & 
\colhead{$\Delta$T} &
\colhead{Survey}   \\
\colhead{} & \colhead{} & \colhead{} & \colhead{[mag]} & \colhead{[pc]} & \colhead{} &
\colhead{[${\rm m\,s}^{-1}$]} & \colhead{[${\rm m\,s}^{-1}$]} & \colhead{[days]} &
\colhead{} 
}
\startdata
GJ 1\tablenotemark{(1)} &439 &M1.5~V & 8.57&4.36 & 15 & 2.6   & 2.6   & 1743   & VLT      \\
GJ 2    &428&M2~V 	&  9.93 &11.5& 9	& 5.2	& 5.1	& 721	& HET 		\\
GJ 15 A\tablenotemark{(2)}&1475&M2~V	&  8.08  &3.6& 29    & 5.7   & 4.9& 2680  & HJS 2.7~m\\ 
GJ 1009 &1734&M1.5~V & 11.16 &18.2  & 11       & 6.1   & 3.3   & 824   & VLT          \\
GJ 27.1 &3143&M0.5~V & 11.42 &22.9  & 18       & 6.7   & 4.7   & 1443   & VLT          \\
GJ 38	&4012&M2~V	& 10.67	 &18.4& 9	& 6.3 	& 6.5	& 487	& HET		\\
GJ 87   &10279&M2.5~V	& 10.06  &10.4& 17	& 8.9 	& 7.3	& 1209 	& HET		\\
GJ 96	&11048&M1.5~V	&  9.41	 &11.9& 17	& 6.8	& 5.1	& 386 	& HET		\\
GJ 109	&12781&M3.5~V	& 10.57	 &7.6& 6	& 6.4	& 11.6	& 388	& HET		\\
GJ 118  &13389&M2.5~V   & 10.7  & 11.5 &  19    & 5.1   & 5.9  & 1769   & VLT          \\
GJ 155.1 &17743&M1~V 	& 11.04  &17.4& 6	& 7.1	& 16.8	& 770	& HET		\\
GJ 160.2 &19165 &M0~V & 9.69 & 23.5 & 14       & 7.9   & 6.3   & 1451   & VLT          \\
GJ 162	&19337&M1~V	& 10.18	 &13.7& 8	& 7.1	& 7.7	& 381	& HET		\\
GJ 176	&21932&M2.5~V	&  9.98	 &9.4& 10	& 5.8	& 7.8	& 420	& HET		\\
GJ 179	&22627&M4~V	& 11.98	 &12.1& 9	& 13.9	& 18.6	& 422	& HET		\\
GJ 180  &22762 & M2~V   & 12.5  & 12.4 & 13       & 3.7   & 3.0   & 1453   & VLT         \\
GJ 181	&23147& M2~V	&  9.78	 &16.5& 7	& 7.4	& 4.9	& 472	& HET		\\
GJ 184  &23518& M0~V 	&  9.93  &14.0& 8 	& 2.8  	& 4.3	& 796	& HET		\\
GJ 192  &24284& M3.5~V& 10.76  &12.7& 10	& 9.4  	& 11.8	& 448	& HET		\\
GJ 3352 &26113& M3~V	& 11.07  &26.6& 14	& 10.9	& 13.1	& 772   & HET		\\ 
GJ 229 A &29295 & M2~V & 8.14 &5.77 & 22       & 4.9   & 3.0  & 1780   & VLT         \\
GJ 251.1 &33241& M1.5~V& 10.55  &48.5& 18 	& 9.1 	& 10.1	& 674 	& HET	\\
GJ 270\tablenotemark{(2)}&35495&M0~V& 10.07&19.8&28& 12.0& 11.0	& 1120	& HET \\
GJ 272  & 35821& M2~V 	& 10.53  &16.2& 22  	& 9.5  	& 11.0	& 1099 	& HET		\\
GJ 277.1 &36834& M0~V 	& 10.49	 &11.5& 14	& 7.2  	& 13.7	& 403 	& HET		\\
GJ 281   &37288& M0~V 	&  9.61  &14.9& 11 	& 8.9 	& 5.0	& 734	& HET		\\
GJ 289   &38082& M2~V 	& 11.46  &14.1& 11 	& 8.2 	& 10.3	& 413	& HET		\\
GJ 308.1 &41689& M0~V 	& 10.33  &19.1& 33	& 10.5	& 10.3	& 716 	& HET	 	\\
GJ 310\tablenotemark{(2)}&42220& M1~V 	&  9.30  &13.9& 38	& 13.7  & 4.1	& 1104  & HET\\
GJ 328   &43790& M1~V 	&  9.99	 &20.0& 11 	& 11.6 	& 5.2	& 442  	& HET		\\
GJ 353   &46769& M2~V 	& 10.19  &13.5& 11 	& 7.5 	& 15.1	& 408 	& HET		\\
GJ 357   &47103 & M2.5~V & 10.85 & 8.98 & 18      & 3.7   & 2.5  & 1221   & VLT         \\
GJ 378   &49189& M2~V 	& 10.07  &14.9&  6 	& 5.2  	& 5.7	& 747 	& HET		\\
GJ 411\tablenotemark{(1)}&54035& M2~V	&  7.48  &2.5&  24& 5.6 & 4.5 	& 2536 	& HJS 2.7~m\\
GJ 412 A &54211& M2~V	&  8.68	 &4.8&  23   & 6.9	& 7.1	& 1826	& HJS 2.7~m	\\ 
GJ 2085	&55625& M1~V	& 11.18	 &21.2&  13	& 12.1	& 17.9	& 407	& HET		\\
GJ 430.1 &56238& M1~V	& 10.30	 &16.2&  21	& 10.8	& 11.7	& 460	& HET		\\
GJ 433   &56528& M1.5~V& 9.79 & 9.04 & 41   &   4.3  &  3.6 &    1938  & VLT \\
GJ 436\tablenotemark{(3)}&57087& M2.5~V& 10.67  &10.2&  57  & 16.6  & 14.5 & 1245 	& HET \\
Wolf 9381 &58114& M1.5~V& 11.50	 &27.9&  14	& 13.6	& 18.1	& 386	& HET		\\
GJ 1170	&64880 &M2~V	& 11.29	 &21.6&  7	& 6.8 	& 13.2 	& 353	& HET		\\
GJ 510   &65520& M1~V   &  11.05  &16.3&  15      & 5.1   & 3.7   & 423   & VLT          \\
GJ 535   &68337& M0~V 	&  9.03  &23.8&  8	& 6.8	& 3.0	& 769 	& HET		\\
GJ 551\tablenotemark{(1)} &70890& M5.5~V& 11.05  & 1.29 & 69     & 3.6   & 2.3   & 1886   & VLT       \\ 
GJ 552   &70865& M2.5~V& 10.68  &14.3&  9	& 4.4  	& 6.2	& 769 	& HET		\\
GJ 563.1 &72387& M2~V 	&  9.71  &24.2&  10	& 10.9  & 8.0 	& 767 	& HET		\\
GJ 618.1 &80053& M2~V	& 10.70	 &30.3& 7	& 6.8	& 9.6	& 440	& HET		\\
GJ 623\tablenotemark{(4)}&80346& M3~V 	& 10.28  &8.0& 22  & 7.7 & 10.6	& 892 	& HET	\\
GJ 637  &82256& M0.5~V    & 11.36  &15.9 & 17 & 6.4  & 3.8  & 1099  & VLT          \\
GJ 655	&83762& M3~V	& 11.61  &13.5& 40 & 13.4	& 22.3	& 1178	& HET		\\
GJ 2128	&84521& M3.5~V& 11.49  &14.9& 9	& 7.4	& 15.9	& 736   & HET	 	\\
GJ 671   &84790& M3~V 	& 11.37  &12.3& 12 	& 10.1 	& 14.1	& 742	& HET	\\
GJ 682   &86214& M3.5~V   & 10.96  &5.04 & 17 & 4.3  & 2.4  & 1050   & VLT        \\
GJ 687   &86162& M3.5~V&  9.18  &4.5& 25 	& 9.9	& 6.5	& 2514	& HJS 2.7~m		\\
GJ 699\tablenotemark{(1)}&87937& M4~V 	&  9.53  &1.8& 70& 3.4 & 2.7	& 1967	& VLT	\\
GJ 709	&89560& M0~V	& 10.28	 &17.1& 8	&  8.8	& 12.5	& 427	& HET		\\
GJ 4070	&91699& M3~V	& 11.27	 &11.3& 10	&  6.6	& 16.4	& 822	& HET		\\
GJ 725 A\tablenotemark{(2)}&91768& M3~V & 8.91  & 3.57 & 23 &  7.4  & 8.1  & 2588   & HJS 2.7~m        \\
GJ 725 B\tablenotemark{(2)}&91772 & M3.5~V & 9.69 & 3.52 & 20 &  7.1  & 9.4  & 2588   & HJS 2.7~m        \\
GJ 730	&92417& M1.5~V& 10.74	 &21.8& 18	& 13.3	& 12.7	& 822	& HET		\\
GJ 731   &92573& M1.5~V& 10.15  &15.6& 8 	&  8.2  & 5.9	& 852	& HET		\\
GJ 739   &93206 & M2~V& 11.14 &13.9 & 19    &  4.6  & 3.3   & 1062   & VLT        \\
GJ 817 &104059& M1~V  & 11.48  &19.1& 20   & 4.9  & 4.0  & 1139   & VLT         \\
GJ 821 &104432& M1~V  & 10.87  &12.1& 31   & 5.3  & 3.8  & 1155   & VLT          \\
GJ 828.1 &105885& M1~V	& 10.52	 &28.6& 9	& 11.3	& 13.4	& 425	& HET		\\
GJ 839   &108092& M1~V 	& 10.35  &23.2& 11 	&  9.0	& 6.9	& 722	& HET		\\
GJ 846	 &108782& M0.5~V& 9.18	 &10.3& 6	&  2.9	& 4.2	& 352	& HET		\\
GJ 849   &109388& M3.5~V& 10.37  &8.8 & 7 	&  8.8 	& 10.3	& 657	& HET		\\
GJ 855	 &110534 & M0.5~V& 10.74  &19.4 & 17 & 5.7  & 4.6  & 1152   & VLT         \\
GJ 864\tablenotemark{(2)}&111571& M1~V 	& 10.01  &17.5 & 27 	& 10.4  & 10.7	& 999 	& HET	\\
GJ 891   &114411& M2~V  & 12.2 & 15.7 & 16& 5.5 & 4.0   & 1768  & VLT          \\
GJ 894.1 &115058& M0.5~V& 10.90	 &24.3& 7	&  8.4	& 13.0	& 361	& HET		\\
GJ 895   &115562& M2~V	& 10.04  &13.1 & 10 	&  9.2  & 6.9	& 788 	& HET		\\
GJ 899	& 116317 & M4~V	& 11.17	 &14.0& 6	&  4.0	& 12.2	& 333	& HET		\\
GJ 911  & 117886 & M0~V & 10.88  &24.7& 11 & 7.8  & 5.4  & 834   & VLT        \\
BD+07 499 & 15563     & M0~V	 &  9.77 &29.3& 12	& 18.4	& 5.1	& 2291	& KECK  	\\
HD 285590 & 19862     & M2~V	 & 11.20&  32.1 &  9	&  7.0  & 4.8	& 2291  & KECK  	\\  
HD 286363 & 18322     & M0~V	& 10.15  & 37.8 & 10	& 13.4	& 4.9	& 1863	& KECK  	\\  
HD 286554 & 19316     & M0~V	& 11.37	 & 40.2&  9    & 18.7	& 7.0	& 1864	& KECK  	\\  
HIP 15720 & 15720     & M0~V   & 11.12 & 33.6& 10	&  8.8	& 7.5	& 1896	& KECK  	\\ 
HIP 16548 & 16548     & M0~V   & 12.01 & 58.1&  9	& 17.5	& 9.5	& 1864	& KECK  	\\
HIP 17766 & 17766     & M1~V	 & 10.93 &41.6&  9	&  6.4	& 6.5	& 1751	& KECK  	\\
vA 115\tablenotemark{(5)}&  ...... & M1~V 	 & 12.98&...&  9   	& 19.3	& 12.9	& 1864	& KECK  	\\
vA 146\tablenotemark{(5)}&  ...... & M1~V	 & 11.98&...&  7	&  8.5	& 7.0	& 1864	& KECK  	\\
vA 383\tablenotemark{(5)}& ......  & M1~V	& 12.24	 &...&  6 	& 11.4	& 6.9	& 1081	& KECK  	\\
vA 502\tablenotemark{(5)}& ......  & M1~V	& 11.95	 &...&  9	& 20.2	& 10.4	& 1836	& KECK  	\\            
vA 731\tablenotemark{(5)}&......& M0~V	& 12.83	 &...&  5	& 5.6	& 6.4	& 1081	& KECK  	\\   
Melotte 25 303\tablenotemark{(5)}&......	& M0~V	 & 11.51&...&  6	& 5.8	& 5.4	& 1398	& KECK		\\
Melotte 25 332\tablenotemark{(5)}&......	& M0~V	 & 11.53&...&  7	& 10.2	& 6.8	& 1721	& KECK		\\
Melotte 25 348\tablenotemark{(5)}&......	& M0~V	 & 11.39&...&  5	& 5.2	& 5.4	& 1512	& KECK	\\
\enddata
\tablenotetext{1}{secular acceleration of the RV subtracted, see Appendix}
\tablenotetext{2}{linear RV trend subtracted, see Appendix}
\tablenotetext{3}{GJ~436: $P = 2.64$ days ; Butler et al.~(2004)}
\tablenotetext{4}{GJ~623: binary orbit subtracted, see Appendix}
\tablenotetext{5}{No $Hipparcos$ data}
\end{deluxetable}
\clearpage
\clearpage
\newpage

\begin{thebibliography}{}

\bibitem[2005]{jacob}
Bean, J.L., Benedict, G.F., Endl, M., \& Sneden, C. 2005, AAS, 207, 6818

\bibitem[1999]{fritz1999}
Benedict, G.F., McArthur, B., Chappell, D.W., Nelan, E., et al. 1999, \aj, 118, 1068

\bibitem[2002]{fritz2002}
Benedict, G.F., McArthur, B.E., Forveille, T., Delfosse, X., et al. 2002, \apj, 581, L115 

\bibitem[2004]{bond}
Bond, I.A., Udalski, A., Jaroszynski, M., et al. 2004, \aap, 606, L155 

\bibitem[2005a]{bonfils_a}
Bonfils, X., Delfosse, X., Udry, S., Santos, N.C., et al. 2005a, \aap, 442, 635

\bibitem[2005b]{bonfils_b}
Bonfils, X., Forveille, T., Delfosse, X., Udry, S., et al. 2005b, \aap, 443, L15

\bibitem[2003]{borucki}
Borucki, W.J., Koch, D.G., Basri, G.B., Caldwell, D.A., et al. 2003, ASP Conf.Ser., 294, 427

\bibitem[2006]{boss}
Boss, A. P. 2006, \apj, accepted

\bibitem[2003]{burgasser}
Burgasser, A.J., Kirkpatrick, J.D., Reid, N.I., Brown, M.E., Miskey, C.L., \& Gizis, J.E. 2003, \apj, 586, 512

\bibitem[2004]{butler2004}
Butler, R.P., Vogt, S.S., Marcy, G.W., et al. 2004, \apj, 617, 580

\bibitem[1997]{bill1997}
Cochran, W.D., Hatzes, A.P., Butler, R.P., \& Marcy, G.W. 1997, \apj, 483, 457 

\bibitem[2002]{bill2002}
Cochran, W.D., Hatzes, A.P., \& Paulson, D.B. 2002, \aj, 124, 565

\bibitem[1998]{delfosse}
Delfosse, X., Forveille, T., Mayor, M., Perrier, C., Naef, D., \& Queloz, D. 1998, \aap, 338, L67 

\bibitem[2003]{endl2003}
Endl, M., Cochran, W.D., Tull, R.G., \& MacQueen, P.J. 2003, \aj, 126, 3099

\bibitem[2004]{endl2004}
Endl, M., Hatzes, A.P., Cochran, W.D., McArthur, B., et al. 2004, \apj, 611, 1121

\bibitem[2005]{fischer2005}
Fischer, D.A., \& Valenti, J.A. 2005, \apj, 622, 1102 

\bibitem[2002]{gaudi2002}
Gaudi, B.S., Albrow, M.D., An, J., Beaulieu, J.-P., et al. 2002, \apj, 566 463

\bibitem[1991]{gliese}
Gliese, W., \& Jahrei\ss, H. 1991, Preliminary Version of the Third Catalogue of Nearby Stars,
Astronomical Data Center CD-ROM

\bibitem[2003]{artie2003}
Hatzes, A.P., Cochran, W.D., Endl, M., McArthur, B., et al. 2003, \apj, 599, 1383 

\bibitem[1994]{heintz}
Heintz, W.D., \& Cantor, B.A. 1994, PASP, 106, 363

\bibitem[1998]{henry}  
Henry, T.J. 1998, ASP Conf. Proc. Vol. 134, 28

\bibitem[1999]{huensch}
H\"unsch, M., Schmitt, J.H.M.M., Sterzik, M.F., \& Voges, W. 1999, A\&ASS, 135, 319

\bibitem[2005]{ida}
Ida, S., \& Lin, D.N.C. 2005, \apj, 626, 1045

\bibitem[2005]{konacki}
Konacki, M. 2005, Nature, 436, 230

\bibitem[2004]{martin2004}
K\"urster, M., \& Endl, M. 2004, ASP Conf. Proc. Vol. 321, 84 

\bibitem[2003]{martin2003}
K\"urster, M., Endl, M., Rouesnel, F., Els, S. et al. 2003, \aap, 403, 1077

\bibitem[2004]{laughlin}
Laughlin, G., Bodenheimer, P., \& Adams, F.C. 2004, \apj, 612, L73 

\bibitem[1996]{lin}
Lin, D.N.C., Bodenheimer, P., \& Richardson, D.C. 1996, Nature, 380, 606

\bibitem[2003]{lineweaver}
Lineweaver, C.H., \& Grether, D. 2003, \apj, 598, 1350

\bibitem[1978]{lippincott}
Lippincott, S. L., \& Borgman, E. R. 1978, PASP, 90, 226

\bibitem[2005]{marcy3}
Marcy, G.W., Butler, R.P., Fischer, D., Vogt, S.S., Wright, J.T., Tinney, C.G. \&
Jones, H.R.A. 2005, PThPS, 158, 24

\bibitem[2001]{marcy2}
Marcy, G.W., Butler, R.P., Fischer, D., Vogt, S.S., Lissauer, J.J, \& Rivera, E.J. 2001, \apj, 556, 296

\bibitem[1998]{marcy1}
Marcy, G.W., Butler, R.P., Vogt, S.S., Fischer, D., \& Lissauer, J.J 1998, \apj, 505, L147

\bibitem[1995]{mayor}
Mayor, M., \& Queloz, D., 1995, Nature, 378, 355

\bibitem[2004]{mcarthur2004}
McArthur, B.E., Endl, M., Cochran, W.D., Benedict, G.F., et al. 2004, \apj, 614, L81

\bibitem[2004]{mccarthy2004}
McCarthy, C., \& Zuckerman, B. 2004, \aj, 127, 2871

\bibitem[2002]{nidever}
Nidever, D. L., Marcy, G. W., Butler, R. P., Fischer, D. A., \& Vogt, S. S. 2002, ApJS, 141, 503

\bibitem[2003]{diane03}
Paulson, D.B., Sneden, C., \& Cochran, W.D. 2003, \aj, 125, 3185 

\bibitem[2004]{diane04}
Paulson, D.B., Cochran, W.D., \& Hatzes, A.P. 2004, \aj, 127, 3579

\bibitem[1997]{hipparcos}
Perryman, M. A. C., ed. 1997, The Hipparcos and Tycho Catalogues (ESA SP-1200; Noordwijk: ESA)

\bibitem[1996]{pollack}
Pollack, J.B., Hubickyj, O., Bodenheimer, P., Lissauer, J.J., Podolak, M., \& Greenzweig, Y. 1996,
Icarus, 124, 62

\bibitem[2005]{rivera}
Rivera, E., Lissauer, J., Butler, R.P., Marcy, G.W., Vogt, S., Fischer, D.A., Brown, T., \& Laughlin, G. 2005,
\apj, 634, 625

\bibitem[2004]{santos}
Santos, N.C., Israelian, G., \& Mayor, M. 2004, \aap, 415, 1153

\bibitem[1992]{tokovinin}
Tokovinin, A.A. 1992, \aap, 256, 121

\bibitem[1998]{valenti}
Valenti, J.A., Piskunov, N., \& Johns-Krull, C.M. 1998, \apj, 498, 851

\bibitem[1996]{wetherill}
Wetherill, G. W. 1996, Icarus, 119, 219

\bibitem[2005]{woolf}
Woolf, V. M., \& Wallerstein, G. 2005, MNRAS, 356, 963

\end{thebibliography}
\end{document}